\documentclass[epl, reprint, aps, amsmath, amssymb, superscriptaddress]{revtex4-2}
\usepackage{amsmath}
\usepackage{siunitx}
\usepackage{verbatim}
\usepackage{bm}

\usepackage{lmodern} 
\usepackage[bookmarks, colorlinks=false]{hyperref}
\usepackage{mathtools}
\hypersetup{colorlinks=true, allcolors=blue}

\begin{document}

\newcommand*{\horz}{\rule[.5ex]{2.5ex}{0.5pt}}
\newcommand{\redcomment}[1]{\textcolor{red}{#1}}
\newcommand{\half}{\ensuremath{^{\scriptstyle{1}}\!\!/\!_{\scriptstyle{2}}}}

\title{Simulated TEM imaging of a heavily irradiated metal}

\author{Daniel R. Mason}
	\email{daniel.mason@ukaea.uk}
	\affiliation{UK Atomic Energy Authority, Culham Centre for Fusion Energy, Oxfordshire OX14 3DB, United Kingdom}
	
\author{Max Boleininger}
	\email{max.boleininger@ukaea.uk}
	\affiliation{UK Atomic Energy Authority, Culham Centre for Fusion Energy, Oxfordshire OX14 3DB, United Kingdom}

\author{Jack Haley}
	\email{jack.haley@ukaea.uk}
	\affiliation{UK Atomic Energy Authority, Culham Centre for Fusion Energy, Oxfordshire OX14 3DB, United Kingdom}

\author{Eric Prestat}
	\email{eric.prestat@ukaea.uk}
	\affiliation{UK Atomic Energy Authority, Culham Centre for Fusion Energy, Oxfordshire OX14 3DB, United Kingdom}

\author{Guanze He}
	\email{guanze.he@materials.ox.ac.uk}
	\affiliation{Department of Materials, University of Oxford, Parks Road, Oxford, OX1 3PH, United Kingdom}
	
\author{Felix Hofmann}
	\email{felix.hofmann@eng.ox.ac.uk}
	\affiliation{Department of Engineering, University of Oxford, Parks Road, Oxford, OX1 3PJ, United Kingdom}
	
	
\author{Sergei L. Dudarev}
	\email{sergei.dudarev@ukaea.uk}
	\affiliation{UK Atomic Energy Authority, Culham Centre for Fusion Energy, Oxfordshire OX14 3DB, United Kingdom}

\date{\today}
\begin{abstract}
We recast the Howie-Whelan equations for generating simulated transmission electron microscope (TEM) images, replacing the dependence on local atomic displacements with atomic positions only.
This allows very rapid computation of simulated TEM images for arbitrarily complex atomistic configurations of lattice defects and dislocations in the dynamical two beam approximation.
Large-scale massively-overlapping cascade simulations performed with molecular dynamics, are used to generate representative high-dose nanoscale irradiation damage in tungsten at room temperature, and we compare the simulated TEM images to experimental TEM images with similar irradiation and imaging conditions.
The simulated TEM shows `white-dot' damage in weak-beam dark-field imaging conditions, in line with our experimental observations and as expected from previous studies, and in bright-field conditions a dislocation network is observed.
In this work we can also compare the images to the nanoscale lattice defects in the original atomic structures, and find that at high dose the white spots are not only created by small dislocation loops, but rather arise from nanoscale fluctuations in strains around curved sections of dislocation lines.

\end{abstract}
\pacs{}
\maketitle

\section{Introduction}

The thermomechanical properties of structural materials are strongly dependent on the material microstructure, yet materials proposed for advanced nuclear fission and fusion must tolerate exposure to irradiation, which introduces nanoscale defects changing the microstructure \emph{in service} \cite{You2016,Hasegawa_JNM2011}.
There is, therefore, a real interest in characterizing the evolution of microstructure as a function of irradiation dose and temperature.
A popular and long-established tool for investigating nanoscale defects characteristic of irradiated microstructures is conventional TEM, as it offers a direct window onto the microstructure at the micrometre scale with sub-nanometre resolution\cite{kirk_JMatRes_2015,Jenkins2001}.
Many features of irradiated microstructure can be identified unambiguously with TEM-- voids\cite{Allen_JNM2006,ElAtwani_Acta2018}, bubbles\cite{ElAtwani_SciRep_2014,Elatwani_JNM2015}, large dislocation loops\cite{Yao_JNM2013,Haley_ActaMat2017}, dislocation lines\cite{Hirsch_PhilMagA1956,Wu_JMicroscopy_2019}, stacking faults\cite{Hirsch1967}, stacking fault tetrahedra\cite{Silcox_PhilMagA_1959,Loretto_ScriptaMat2015} and second-phase particles\cite{Baluc2011}.
But small features of a few nanometres in size 
are much more difficult to characterize.
Notwithstanding the issue that a small cluster of defected atoms may image too faintly to be readily detected, by eye or otherwise, those small features which do image brightly typically appear as featureless spots, offering no real insight into their true atomic nature. 
These small features, characteristic of radiation damage, are often referred to as `black-dot' damage when observed in bright-field conditions (or `white-dot' in dark-field)\cite{Yao_PhilMag2008,Prokhodtseva_ActaMat2013,Yi_ActaMat2016}.
The small features are of fundamental importance to understanding microstructural evolution, for as well as being the characteristic state for low temperature irradiation, quasi-independent prismatic dislocation loops are the first defects formed during irradiation\cite{Sand_EPL2013,Yi_EPL2015}, and they also form the building blocks for defect annealing and coarsening.

Image simulations suggest a single, isolated black-dot is consistent with a single, perfect, isolated dislocation loop.
Experimentally, we sometimes find dots that become invisible at a particular set of imaging $\bf{g}$-vectors, suggesting that $\bf{g}\cdot\bf{b}=0$ in these conditions, and hence indicating a loop with a single Burgers vector\cite{Jenkins2001}.
But this invisibility criterion is only really true in the limit of long-range linear elasticity, and cannot be relied on in cases where the defect is much smaller than this elastic limit\cite{Jenkins2001,
Mason_JAP2019}.
This identification is further complicated by the fact that a single, perfect, isolated dislocation loop is often the one thing the spot cannot possibly be:  
molecular dynamics simulations show that small dislocation loops in Body-Centre Cubic (BCC) metals are extremely mobile, with a diffusion constant order 0.01 {\textmu m$^2$/s}\cite{Derlet_PRB2011}, and so prismatic loops would find their way to the TEM foil surface before observation is possible\cite{Schaublin2012}.
This contradiction has been explained previously by assuming the loop is pinned by impurity atoms\cite{Arakawa_Science2007,Castin_JNM2019}, or by elastic interactions with other lattice defects\cite{Mason_JPCM2014}.
Molecular dynamics simulations of overlapping cascades have offered another possibility-- that the defects are not simple loops with one Burgers vector, but may be complex objects\cite{Sand_JNM2018,Byggmaestar_JPCM_2019}.
High-dose, massively overlapping cascade simulations have suggested that the transition from simple loop-like defects to complex network dislocation microstructure may happen at a dose of 0.01 to 0.1 dpa \cite{Derlet_PRM2020,Mason_PRL2020,Warwick_PRM2021}.
The true nature of the black dots is therefore uncertain.

Rather than adopting the standard approach of simulating TEM images of known, isolated defects, and checking they image in a way consistent with experimental observations\cite{Head_Book1973}, in this work we perform image simulations of characteristic irradiated microstructure and view them dispassionately.
While this approach is similar in motivation to that employed by Sch\"aublin {\it et al.} previously\cite{Schaeublin_JNM2002}; our work has important computational differences allowing us to explore larger simulation sizes with a wider range of defect environments.
In section \ref{sec:derivation2BDI}, we rederive the equations for simulating dynamical two-beam imaging. 
We recast the partial differential equations, replacing the dependence on an atomistic displacement field with a dependence only on atomic {\it position}.
This is more physically meaningful for arbitrary atomic configurations typically generated using molecular dynamics simulations where the reference lattice may be difficult to define, and is computationally very efficient, generating a simulated image of a million atom configuration in seconds.

We emphasize that more sophisticated (and therefore more computationally expensive) models for generating simulated TEM or STEM images exist. \texttt{ABTEM}~\cite{abtem} and \texttt{PRISTMATIC}~\cite{RangelDaCosta_Micro2021} consider the dispersion of electrons during their propagation through the material, requiring a multi-slice simulation technique. While more accurate, this added level restricts the number of atoms which can be handled on a desktop computer, and requires more user input in the form of testing for convergence in the solution.
 
In section \ref{sec:MD} we describe large-scale MD simulations used to produce characteristic high-dose irradiated microstructures.
Similar simulations have been shown previously to have a maximum hydrogen retention capacity in agreement with hydrogen plasma-loading experiments\cite{Mason_PRM2021}, a qualitative strain response in agreement with X-ray diffraction measurements\cite{Mason_PRL2020}, ion-beam mixing in agreement with Rutherford Backscattering (channelling) experiments\cite{Gra15,Vel17,Gra20}, and thermal conductivity in agreement with transient grating spectroscopy experiments\cite{Mason_PRM2021b}.
For this work we use simulation boxes of 21 M atoms; from a computational perspective, this size is necessary to minimise elastic periodic image effects from the defects generated, but more importantly the cell side length (70 nm) is directly comparable to the typical thickness of a TEM transparent foil, and the simulation boxes contain sufficient defects to enable a non-trivial statistical image analysis.
An image calculation of this kind takes a couple of minutes on a desktop computer, and has good parallel scaling (currently to 128 cores) on a computer cluster.

Finally, in section \ref{sec:Expt} we use a JEOL 2100 TEM equipped with a LaB6 source to image tungsten foils irradiated to 1 dpa with 20 MeV self ions at room temperature. These irradiation experiments are, as close as possible, a match to the simulations.
We show that there is a very good qualitative agreement between simulated and experimental TEM images.
But when we look closer at the \emph{known} nanoscale defects in the microstructure responsible for the simulated image, we see a rather weak correlation between the position of brightly imaging spots and the dislocation loops.
Rather we find that, in dark field conditions, the brightly imaging regions are due to any strain fields which fluctuate at the nanometre scale.
These strain fields are generated by all defects, be they simple or complex loops or parts of the dislocation line network.
We conclude that black-dot damage is consistent with a range of complex microstructural features with a characteristic nanometre scale.
And so for characterizing a high dose microstructure, we show by direct simulation that it is necessary to consider both dynamical and weak beam conditions.

\section{Howie-Whelan approximation for generating simulated TEM images}
\label{sec:derivation2BDI}

In this section we rederive the Howie-Whelan equations for dynamical two-beam imaging in a form which allows us to use atomic positions only, rather than a strain or displacement field, computed in the elasticity theory approximation.
We start with the Schr\"odinger equation to describe the propagation of high-energy electrons:
\begin{equation}
    -{\hbar ^2\over 2m}{\partial ^2 \over \partial {\bf r}^2}\Psi ({\bf r})+U({\bf r})\Psi ({\bf r})={\hbar ^2{\bf k}^2\over 2m}\Psi ({\bf r}),\label{Schroedinger}
\end{equation}
where $\Psi ({\bf r})$ is the one-electron wave function of a high-energy electron, as a function of position $\bf{r}$, ${\bf k}$ is the wave vector of incident electrons, and $U({\bf r})$ is the potential energy of interaction between the high-energy electron and the atoms. Although the above equation is taken in the non-relativistic form, in the treatment of electron diffraction the relativistic effects can be accounted for by replacing $m$ with the relativistic electron mass \cite{PengDudarevWhelan}.  

We look for the solution in the form of a sum of propagating and diffracted beams, where the amplitude of each varies slowly as a function of spatial coordinates
\begin{equation}
\Psi ({\bf r})=\Phi _0({\bf r})\exp (i{\bf k}\cdot {\bf r}) +\Phi _{\bf g}({\bf r})\exp[i({\bf k}+{\bf g})\cdot {\bf r}], \label{wave_function}    
\end{equation}
where ${\bf g}$ is a reciprocal lattice vector used for imaging. 

A dark field TEM image of microstructure is given by the intensity distribution $I_{\bf g}(x,y)$ computed as $|\Phi _{\bf g}(x, y, L)|^2$ at the exit surface of the foil at $z=L$.

Substituting (\ref{wave_function}) into (\ref{Schroedinger}), we find
\begin{widetext}
\begin{eqnarray}
    {\hbar ^2 {\bf k}^2\over 2m}\exp (i{\bf k}\cdot{\bf r})\Phi _0({\bf r}) +
    {\hbar ^2 ({\bf k}+{\bf g})^2\over 2m}\exp [i({\bf k}+{\bf g})\cdot{\bf r}]\Phi _{\bf g}({\bf r})&&\nonumber \\
    -i{\hbar ^2\over m}\exp (i{\bf k}\cdot{\bf r})\left({\bf k}\cdot {\partial \over \partial {\bf r}}\right)\Phi _0({\bf r}) -i{\hbar ^2\over m}\exp [i({\bf k}+{\bf g})\cdot{\bf r}]\left(({\bf k}+{\bf g})\cdot {\partial \over \partial {\bf r}}\right)\Phi _{\bf g}({\bf r})&&\nonumber \\
    +U({\bf r})\left\{\Phi _0\exp (i{\bf k}\cdot {\bf r}) +\Phi _{\bf g}\exp[i({\bf k}+{\bf g})\cdot {\bf r}]\right\}&=&{\hbar ^2{\bf k}^2\over 2m} \left\{\Phi _0\exp (i{\bf k}\cdot {\bf r}) +\Phi _{\bf g}\exp[i({\bf k}+{\bf g})\cdot {\bf r}]\right\}\nonumber \\
\end{eqnarray}
\end{widetext}
Separating terms associated with either $\exp (i{\bf k}\cdot {\bf r})$ and   $\exp(i[{\bf k}+{\bf g})\cdot {\bf r}]$, and choosing the direction of the $z$ axis in the direction of ${\bf k}$, where $|{\bf k}| \gg| {\bf g}|$, we arrive at a system of coupled equations for the amplitudes of the transmitted and diffracted beams
\begin{eqnarray}
{\partial \over \partial z}\Phi _0({\bf r})&=&-i{U_{-{\bf g}}\over \hbar v} \exp [i{\bf g}\cdot {\bf u}({\bf r})]\Phi _{\bf g}({\bf r}), \nonumber \\
{\partial \over \partial z}\Phi _{\bf g}({\bf r})&=&-i{\epsilon _{\bf g}\over \hbar v}\Phi _{\bf g}({\bf r}) -i{U_{{\bf g}}\over \hbar v} \exp [-i{\bf g}\cdot {\bf u}({\bf r})]\Phi _0({\bf r}),\nonumber \\ \label{eqn:two_beam1}
\end{eqnarray}
where $v=\hbar k/m$ is the velocity of electrons,  ${\bf u}({\bf r})$ is the field of atomic displacements at ${\bf r}$, and $U_{\bf g}$ is the Fourier component of the periodic potential in an ideal crystal with no atomic distortions, ie
\begin{equation*}
    U({\bf r})=\sum_{\bf h}U_{\bf h}\exp(i{\bf h}\cdot {\bf r}),
\end{equation*}
where summation over ${\bf h}$ is performed over reciprocal lattice vectors. Parameter
\begin{equation*}
    \epsilon_{\bf g} = \frac{\hbar ^2 ({\bf k}+{\bf g})^2 }{2m} - \frac{\hbar ^2 {\bf k}^2}{2m}
\end{equation*}
characterises the deviation of the orientation of the incident electron beam from the exact Bragg condition. In applications, it is often advantageous to use imaging conditions where the magnitude of $\epsilon _{\bf g} $ is substantial, even though this implies that the amplitude of $\Phi _{\bf g}({\bf r})$ is relatively small and the overall intensity of the diffraction image is lower. 

The derivation of equations (\ref{eqn:two_beam1}) assumes that the notion of the Fourier component of the potential $U_{\bf g}$ is still well defined, and the field of displacements ${\bf u}({\bf r})$ varies on the scale much larger than the size of the unit cell. This condition is often satisfied even in the core regions of defects and dislocations \cite{PhilMag2003,Boleininger2018,Boleininger2019}. 

In many cases of practical relevance $U_{{\bf g}}=U_{-{\bf g}}<0$, and it is convenient to define the so-called extinction distance $\xi_{\bf g}$
\begin{equation*}
    \xi_{\bf g}= \frac{\pi \hbar v}{ |U_{\bf g}|},
\end{equation*}
characterising the spatial scale of variation of the solution describing diffraction of high-energy electrons in the crystal. 
At the exact Bragg condition $\epsilon_{\bf g} =0$, and distance $\xi_{\bf g}$ corresponds to a half the period of oscillation of solutions of (\ref{eqn:two_beam1}).

Finally, as a purely technical note, we observe that in literature \cite{Hirsch1967,Arakawa2020}, the energy parameter $\epsilon _{\bf g}$ is often replaced by another parameter $s_{\bf g}$ that also characterizes the deviation of the direction of the incident beam from the exact Bragg diffraction condition. This parameter is defined as 
\begin{equation*}
s_{\bf g}=-\frac{1}{2\pi} \frac{\epsilon_{\bf g}}{\hbar v}.
\end{equation*}
Using the above notation, and also adding the term describing the effect of the average constant crystal potential, equations (\ref{eqn:two_beam1}) can be written in the form \cite{HowieWhelan1961}
\begin{eqnarray}
    \label{eqn:two_beam2}
{\partial \over \partial z}\Phi _0({\bf r})&=&{i \pi \over \xi _{0} }\Phi _0({\bf r})+ {i \pi \over \xi _{\bf g}} \exp [i{\bf g}\cdot {\bf u}({\bf r})]\Phi _{\bf g}({\bf r}), \nonumber \\
{\partial \over \partial z}\Phi _{\bf g}({\bf r})&=&\left({i \pi \over \xi _{0}} + i 2 \pi s_{\bf g} \right) \Phi _{\bf g}({\bf r}) + {i \pi \over \xi_{{\bf g}}} \exp [-i{\bf g}\cdot {\bf u}({\bf r})]\Phi _0({\bf r}).   \nonumber \\ 
\end{eqnarray}
Note that in our notations, the reciprocal lattice vectors already include the factor of $2\pi$ that in earlier literature on electron microscopy is still written separately in the relevant formulae \cite{HowieWhelan1961,Arakawa2020}.

To integrate equations (\ref{eqn:two_beam2}), we need smoothly varying fields of atomic displacements. 
Consider atoms at positions $\mathbf{R}_j$, where $j=\{1,2,\ldots,N\}$. 
If we imagine perfect reference positions for these atoms, $\mathbf{R}^{(0)}_j$, then small displacements $\mathbf{u}_j = \mathbf{R}_j - \mathbf{R}^{(0)}_j$ are well defined. 
But near a dislocation core, displacements are not small, and even the reference lattice site for a given atom may be difficult to uniquely identify.
To solve this problem, note that to integrate equation \ref{eqn:two_beam2} we actually only need the \emph{phase factor} $\exp \left[i \mathbf{g}\cdot \mathbf{u}(\bf r) \right]$.

If our imaging vector $\bf{g}$ is a reciprocal lattice vector satisfying
    \begin{equation}
        \label{eqn:imaging_vector_g}
        \exp \left(i \mathbf{g}\cdot \mathbf{R}^{(0)}_j \right) = 1,
    \end{equation}
we find that at atom positions $\exp \left(i \mathbf{g}\cdot \mathbf{u}_j \right) = \exp \left(i \mathbf{g}\cdot \mathbf{R}_j \right)$.
Therefore, to provide an unambiguous answer for the smoothly varying field of displacements needed to integrate equation \ref{eqn:two_beam2}, we can interpolate the phase factor from atomic {\it positions} themselves, and not the atomic {\it displacements}, by interpolating the value of $\exp \left(i \mathbf{g}\cdot \mathbf{R}_j \right)$,
\begin{equation}
    \label{eqn:interpolate_phase_factor}
    x(\mathbf{r}) = \sum_j \kappa( \mathbf{r} ; \mathbf{R}_j ) \exp \left(i \mathbf{g}\cdot \mathbf{R}_j \right),
\end{equation}
where $\kappa$ is a suitable local interpolation kernel normalised such that $|x(\mathbf{r})| = 1$.
We use the Gaussian form
    \begin{equation*}
        \kappa( \mathbf{r} ; \mathbf{R}_j ) \sim  \exp\left( - \frac{|\mathbf{r}-\mathbf{R}_j|^2}{2 \sigma^2} \right) H\left( 3 \sigma - |\mathbf{r}-\mathbf{R}_j| \right),
    \end{equation*}
where $H(r)$ is the Heaviside function and the lengthscale $\sigma$ is taken to be approximately half the unit cell parameter.

One can then reformulate equations (\ref{eqn:two_beam2}) using a gauge transformation defined by the formulae
\begin{eqnarray}
\Phi_0({\bf r})&=&\phi_0({\bf r}), \nonumber \\
\Phi_{\bf g}({\bf r})&=&x^{-1}(\mathbf{r}) \phi_{\bf g}({\bf r}).
\end{eqnarray}
This does not affect the images since $|\Phi_{\bf g}({\bf r})|^2=|\phi_{\bf g}({\bf r})|^2$. 
As a final detail, we note that in a void region electron waves propagate freely. 
If we write an atomic density function $\rho(\bf{r})$ defined as equal to one in the crystal and zero in void regions, then we find that equations (\ref{eqn:two_beam2}) acquire the final form to be numerically integrated for each pixel in the output image:
\begin{widetext}
\begin{equation}    
    \label{eqn:two_beam3}
{\partial \over \partial z}
\left( \begin{array}{c}
    \phi_0(z) \\
    \phi_{\bf{g}} (z)
\end{array} \right)
=
i \pi \left[ 
\left( \begin{array}{c c}
    0    &   0          \\
    0    & 2 s_{\bf{g}} 
\end{array} \right) 
+
\rho(\bf{r})
\left( \begin{array}{c c}
    1/\xi_0   &   1/\xi_{\bf{g}}      \\
    1/\xi_{\bf{g}}    & 1/\xi_0
\end{array} \right) 
+
\rho(\bf{r})
\left( \begin{array}{c c}
    0    &   0     \\
    0    &  x^{-1}(\mathbf{r}) {\partial \over \partial z}x(\mathbf{r})
\end{array} \right) 
\right]
\left( \begin{array}{c}
    \phi_0(z) \\
    \phi_{\bf{g}} (z)
\end{array} \right) 
\end{equation}
\end{widetext}

As an aside, we note that we can re-express the phase-factor term using 
    $$
        x^{-1}(\mathbf{r}) {\partial \over \partial z}x(\mathbf{r}) = i \mathbf{g} \cdot \left[ \mathbf{F}(\bf{r}) \, \hat{\mathbf{z}} \right],
    $$
where $\mathbf{F}(\bf{r})$ is the deformation gradient tensor~\cite{Roters2010} at position $\bf{r}$ with elements $F_{ij} = \partial {u}_i/ \partial {x}_j$, and  $\hat{\mathbf{z}}$ is a unit vector pointing along $z$.
This shows explicitly that it is the deformation gradient tensor projected onto the imaging vector ${\bf g}$ and the direction of propagation of the electron beam $\hat {\bf z}$, rather than atomic displacements as such, that is responsible for the observed variation of intensity in TEM images of defects. 

We can recognise the three matrix terms in equation \ref{eqn:two_beam3} as the partial differential equation coefficients for propagation through vacuum, propagation through perfect crystal, and the effective change in the local lattice orientation due to the strain field.
Lattice strain and rotation enters the equations, through the lattice deformation gradient tensor, {\it via} the third term in (\ref{eqn:two_beam3}) that is analogous to the deviation $s_{\bf g}$ from the exact Bragg diffraction conditions entering the first term in the same equation. The effect of local lattice distortions on the propagation of high-energy electrons is therefore exactly equivalent to the effect of variation of the local Bragg condition. 

Furthermore, our phase field $x(\mathbf{r})$ has an associated multi-valued coordinate-dependent displacement field, $\mathbf{\cal U}({\bf r})$, defined by $x(\mathbf{r}) = \exp[ i \mathbf{g}\cdot \mathbf{\cal U}(\mathbf{r}) ]$.
Near atom sites, this displacement field satisfies  
    $$
        \mathbf{g}\cdot \mathbf{\cal U}(\mathbf{r})=\mathbf{g}\cdot \mathbf{\cal U}(\mathbf{R}_j + \boldsymbol{\delta} ) = \mathbf{g} \cdot \mathbf{u}_j + \mathbf{g}\cdot \left( \mathbf{F} \, {\boldsymbol{\delta}} \right) + 2 n \pi,    
    $$
where $n$ is an integer. 
This shows that by introducing a single-valued phase field $x(\mathbf{r})$, we are able to truly circumvent the problem of having a multi-valued displacement field near a dislocation core.

We use a fourth-order Runge-Kutta integration scheme with four samples of $x(z)$ per unit cell length.
The extinction distances are computed from the crystal structure factors $|U_{\bf{g}}|$ in the Doyle-Turner approximation~\cite{Dudarev_SurfSci1995,Peng_ActaCrys1996a} including a finite temperature Debye-Waller factor~\cite{Peng_ActaCrys1996b}.

More technical details about our implementation of these equations are given in the Appendix.
We validate the correct functioning of the code by comparing to TEMACI \cite{Zhou_PhilMag_2006} in the supplementary material, section~\ref{sec:validation}.  
Choosing a foil orientation for a good two-beam condition is discussed in section~\ref{sec:two_beam}.
The geometry of how the imaging space relates to the frame represented by the input atomic positions is in section~\ref{sec:image_space}.
 
\section{Generation of high-dose simulated microstructure}
\label{sec:MD}

To be able to compare simulated TEM with ground truth high dose network dislocation microstructures, we need to generate representative atomic configurations.
Recently it has been shown that massively overlapping MD cascade simulations~\cite{Gra15,Byg18,Vel17,Gra20,Granberg_JNM2021} produce high dose microstructures with characteristic properties consistent with several independent experiments: 
Zhang {\it et al.}~\cite{zha17} and Markelj {\it et al.}~\cite{Markelj_ActaMat2023} have shown simulations agree with Rutherford Back Scattering Channeling measurements; Mason {\it et al.} have shown that changes in lattice strain in high dose simulations of tungsten are consistent with micro-Laue X-ray diffraction~\cite{Mason_PRL2020}, the vacancy content in high dose simulations of tungsten is consistent with the observed saturated deuterium retention~\cite{Mason_PRM2021}, and the thermal diffusivity consistent with transient grating spectroscopy experiments~\cite{Mason_PRM2021b}. But these are all calculations of ensemble properties of the simulated microstructures, and here we need to consider the spatial variation in the microstructure.

We therefore use here the simulated irradiation microstructures generated for an earlier paper by some of the authors, Ref~\cite{Boleininger_PRM2022}, where full details of the simulation methodology can be found.
Briefly, a large single crystal box of 21 M tungsten atoms was subjected to multiple 10 keV PKA collision cascades using the MD code \texttt{LAMMPS}~\cite{LAMMPS}, with traction-free periodic boundary conditions, until an NRT dose~\cite{norgett1975proposed} of 1 dpa was reached at order $6 \times 10^5$ cascades total.

These simulation boxes have a good size for our purpose, as the simulation cell side of 70 nm is comparable to TEM foil thickness. Sub-optimally they have periodic boundary conditions in all directions, rather than a free surface, so there are no image forces on the microstructure~\cite{Fikar_NIMB2017}, and no loop loss to the surface~\cite{Zheng_ActaMat2020}.
To generate the images, we take multiple periodic replicas in the plane normal to the beam direction, and a single replica parallel to the beam, essentially creating a new, unrelaxed surface. The geometry of the image generation is discussed in section \ref{sec:image_space}. 

In these and similar massively overlapping cascade simulations in bcc metals, the microstructure at low dose (order 0.01 dpa) shows small, isolated, interstitial character dislocation loops. The size and the Burgers vector of these loops is somewhat potential dependent~\cite{Granberg_JNM2021}-- here we used the potential of Ref~\cite{Mason_JPCM2017}.
At a critical dose (order 0.1 dpa), the density of the loops gets so high that they are no longer separated, but rather form complex networks.
At high dose, the loops have become so large that they span the simulation cell, and can be better described as planes of perfect crystal cut by dislocation lines. At all doses we see some dislocation loops, but we note that MD simulations have a short timescale, order tens of nanoseconds, and so these isolated loops might be able to diffuse and further coalesce if evolved over a longer timescale or at higher temperature.

\section{Experimental imaging of high-dose microstructure}
\label{sec:Expt}

TEM imaging of dislocation loops was carried out on a set of tungsten samples irradiated to 0.01,0.1, and 1.0 dpa. 
These samples were electro-polished and then irradiated by 20 MeV self ions at room temperature.
Images were taken using weak beam dark field conditions on a JEOL-2100 TEM (accelerating voltage 200 kV, LaB6 source).
All images were taken with $\mathbf{g}=[200]$ and zone axis $[001]$ diffraction conditions.
A detailed description can be found in~\cite{He_scripta2023}.

\section{Results and Discussion}

In this section we show images of simulated high-dose microstructure and compare to overlays of the ground-truth dislocations and to experimental TEM images.

\begin{figure*}
    \centering
    \includegraphics[width=0.6\linewidth]{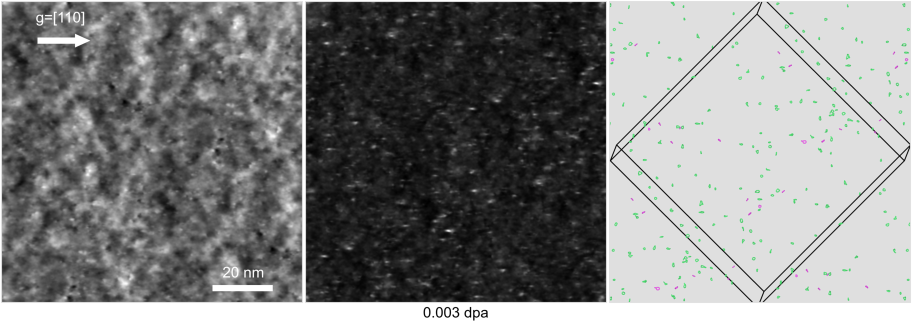}
    \includegraphics[width=0.6\linewidth]{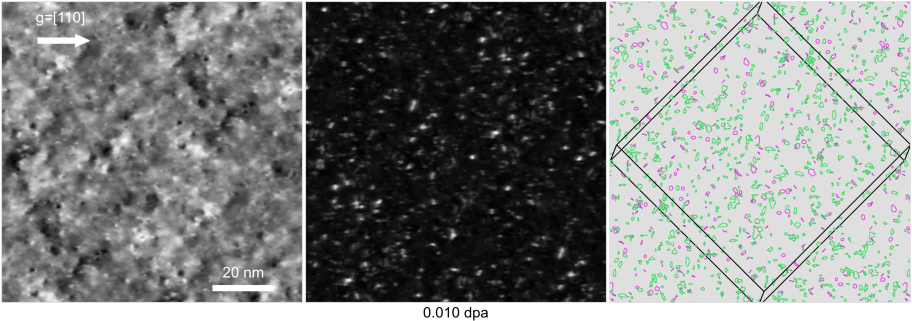}
    \includegraphics[width=0.6\linewidth]{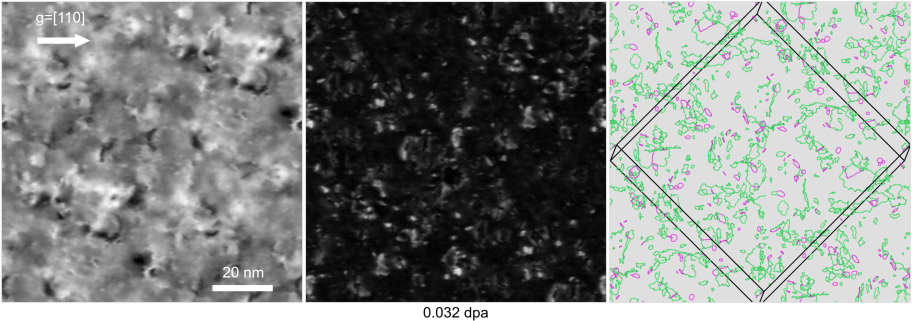}
    \includegraphics[width=0.6\linewidth]{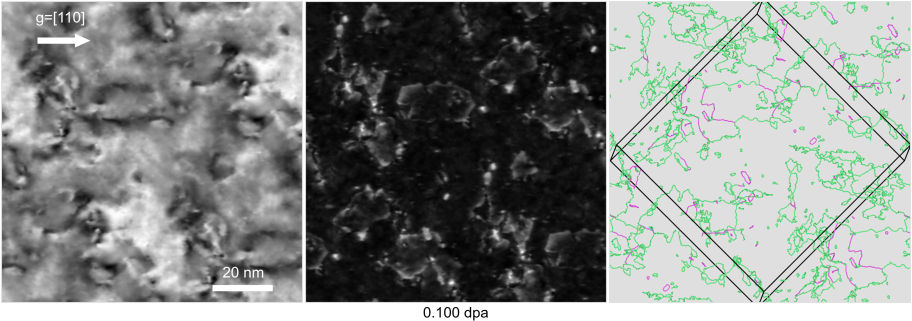}
    \includegraphics[width=0.6\linewidth]{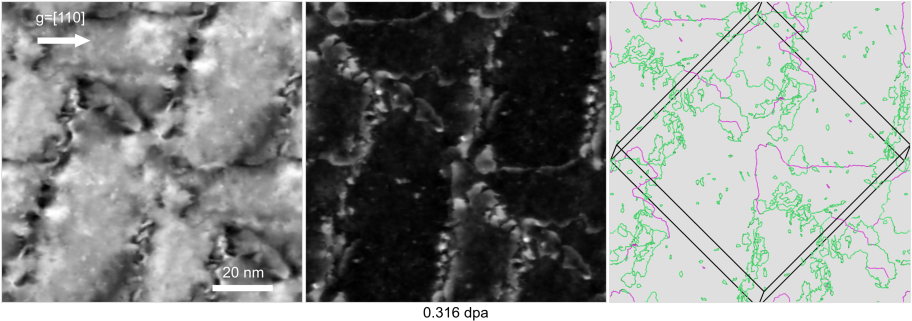}
    \includegraphics[width=0.6\linewidth]{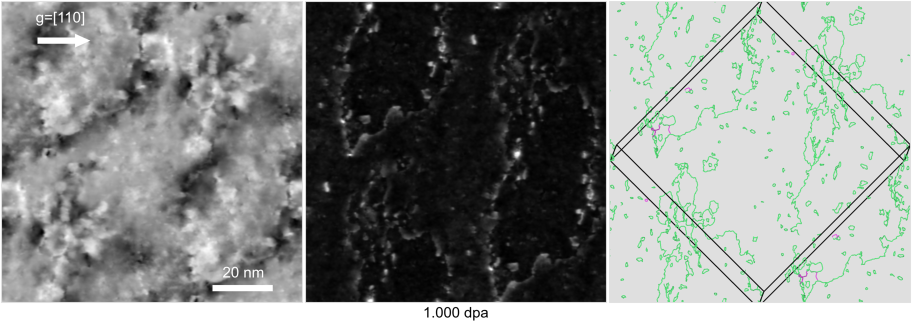}
    \caption{Snapshots of massively overlapping cascade simulations with zone axis close to the [001] direction. At each dose, the three panels are Left: simulated kinematical bright-field two-beam image with $\mathbf{g}=[110]$ horizontal; Centre: weak-beam dark field image; Right: dislocation network shown computed with Ovito\cite{Stukowski_MSMSE2009} ( green lines $1/2\langle 111 \rangle$, pink lines $\langle 100 \rangle$ ).}
    \label{fig:snapshots}
\end{figure*}

\begin{figure}
    \centering
    \includegraphics[width=0.8\linewidth]{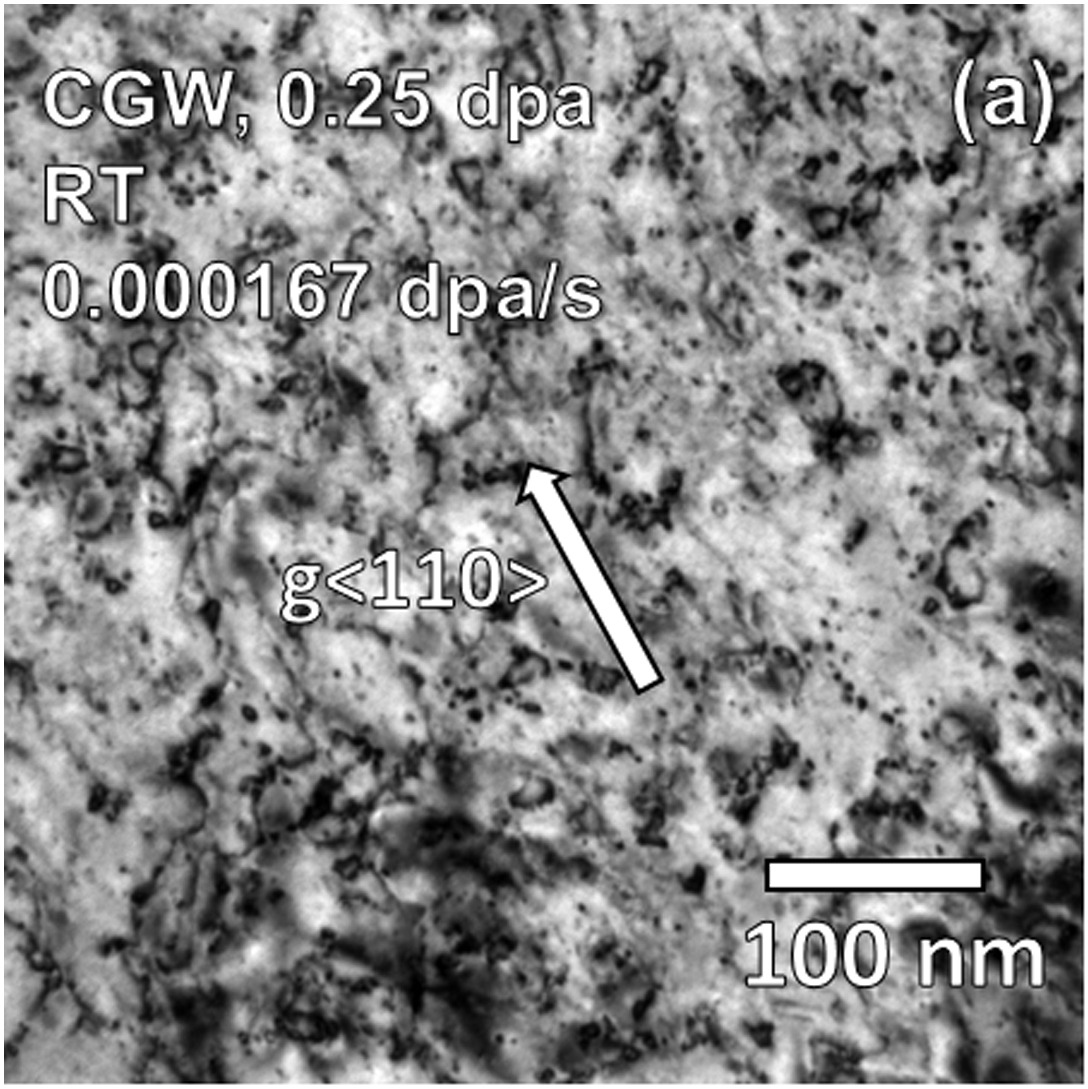}
    \caption{Bright-field experimental image of tungsten, ion-irradiated at room temperature to 0.25 dpa, reproduced from Ref~\cite{ElAtwani_Acta2018a}. Note the simulated images in this work are about 100 nm across, the size of the scale bar in this image.}
    \label{fig:elAtwani}
\end{figure}

The principal result of this work is shown in figure \ref{fig:snapshots}. This shows the evolution of the microstructure from dislocation loops through a network to dislocation lines. Kinematical bright-field and weak-beam dark-field images were generated with $s_{\bf g} \approx 0.03 \rm{nm}^{-1}$ and $s_{\bf g} \approx 0.20 \rm{nm}^{-1}$ respectively, corresponding to $n_{\bf g} = 1.50$ and $n_{\bf g} = 6.25$.
A description of how the foil orientation is chosen to find these deviation parameters is given in appendix~\ref{sec:two_beam}.
In these, and subsequent simulated TEM images, we use an accelerating voltage of 200 keV, and assume a sample temperature of 300K for the extinction distances.
The bright-field images can be compared directly to the images of high-dose, room-temperature tungsten irradiation by El-Atwani {\it et al.}~\cite{ElAtwani_Acta2018,ElAtwani_Acta2018a} who  observed spatially ordered structures at the tens of nanometre lengthscales.
An image from Ref~\cite{ElAtwani_Acta2018a} is reproduced in figure \ref{fig:elAtwani}. 
We see the same qualitative pattern of a high density of curved dislocation lines in both experiment, fig~\ref{fig:elAtwani} and the corresponding dose in simulation, fig~\ref{fig:snapshots}.
El Atwani {\it et al.} attributed the spatial ordering to loop rafting~\cite{Dudarev_JNM2014}, where mobile loops with the same Burgers vector orient and align themselves to reduce their mutual elastic interaction energy.
Here, we have found the complex dislocation network is generated at low temperature in such a way as to lower elastic energy density, which leads to orientation and alignment of parts of the the network without the need for assuming the existence of independent mobile dislocation loops. This is in better agreement with the network growth mechanism suggested by Wang {\it et al.}~\cite{Wang_Acta2023}.
Note that our simulations are low-temperature irradiation, and so would not show the loop rafting mechanism where the defect density is lower at high temperatures.


The image size of the spots seen at low dose in weak beam conditions in figure \ref{fig:snapshots} is very similar to the dislocation loop size. 
But at high dose it is clear that the spot size in weak beam conditions is correlated with the curved sections of dislocation lines. Individual small dislocation loops are present in the high-dose simulation images but image much more faintly in comparison. A discussion of brightly and faintly imaging defects in otherwise perfect crystal is given in section \ref{sec:validation}.

\begin{figure}
    \centering
    \includegraphics[width=0.8\linewidth]{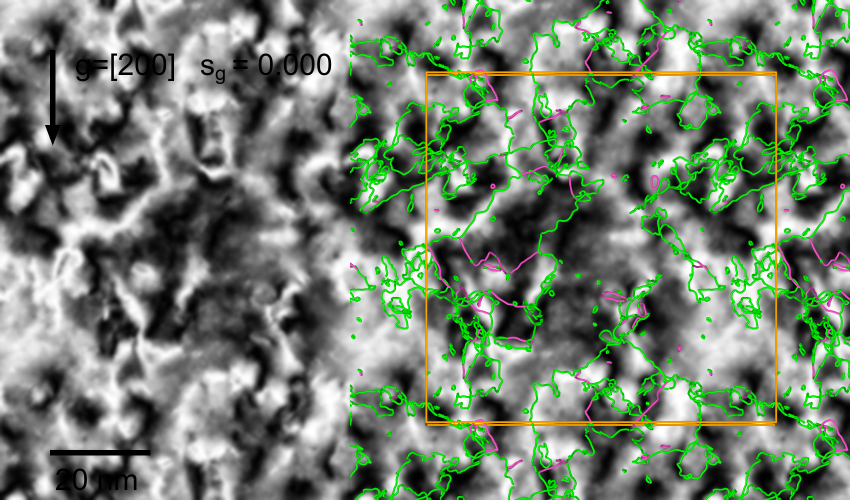} 
    \includegraphics[width=0.8\linewidth]{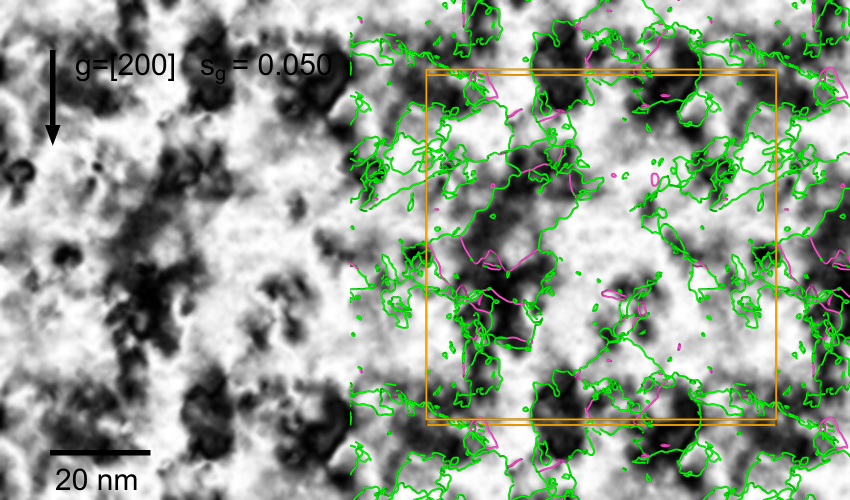} 
    \includegraphics[width=0.8\linewidth]{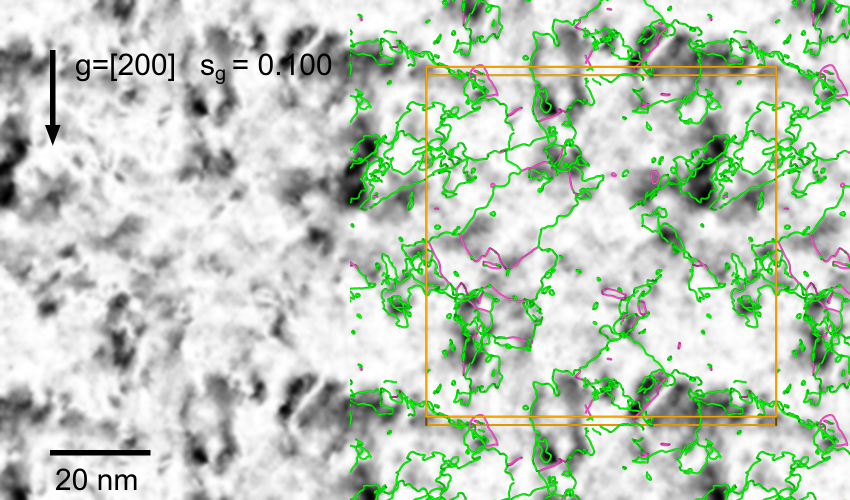} 
    \includegraphics[width=0.8\linewidth]{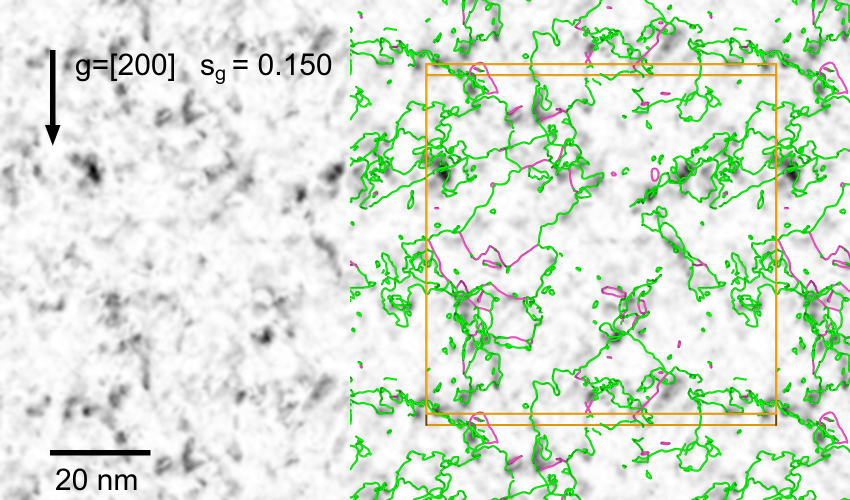}
    \includegraphics[width=0.8\linewidth]{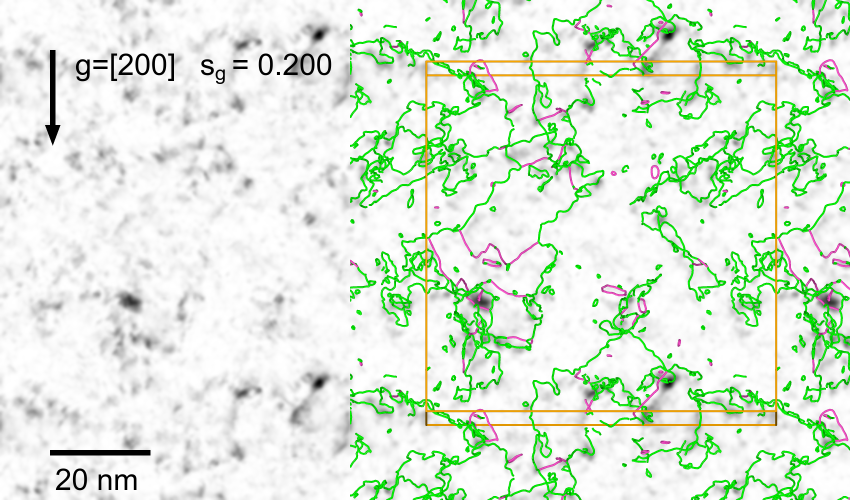}
    \caption{Snapshots of overlapping cascade simulations at 0.1 dpa, with zone axis close to the [001] direction, computed for different deviation parameters $s_g$. Top dynamical two-beam conditions, rotating to bottom weak beam conditions. Dislocation network computed with Ovito\cite{Stukowski_MSMSE2009} overlaid coloured by Burgers vector: green $\half \langle 111 \rangle$ type and pink $\langle 100 \rangle$. All images shown with inverted contrast. Note we have exploited the  simulation box periodic boundary conditions to extend the images.}
    \label{fig:sg}
\end{figure}

To understand the difference in the bright-field and dark-field images, we consider a high-dose microstructure snapshot at a range of deviation parameters. 
Figure \ref{fig:sg} shows a series of images with the same $\mathbf{g} = [200]$ imaging vector and zone axis close to [001], with varying fine tilt in the g-vector direction of the order of a few degrees to change the value of $s_g$. The images go from dynamical two beam condition to weak beam condition.
The dose for this sequence of images is 0.1 dpa, where the microstructure is one of densely packed complex network dislocations and dislocation loops.
From equation \ref{eqn:two_beam3}, we see that in kinematical conditions, $|x^{-1} {\partial x}/\partial z| \gg |s_{\bf g}|$, image intensity associated with the long range elastic fields is seen. 
As the foil is tilted to weak-beam condition, $|s_{\bf g} \gg |x^{-1} {\partial x}/\partial z|$, we only see short range intensity peaks at the positions of the highest strain.
The spatial correlation between the simulated TEM image and the ground truth dislocation configuration is equally good in each image, we simply highlight different magnitudes of the strain field.

\begin{figure*}
    \centering
    \includegraphics[width=0.75\linewidth]{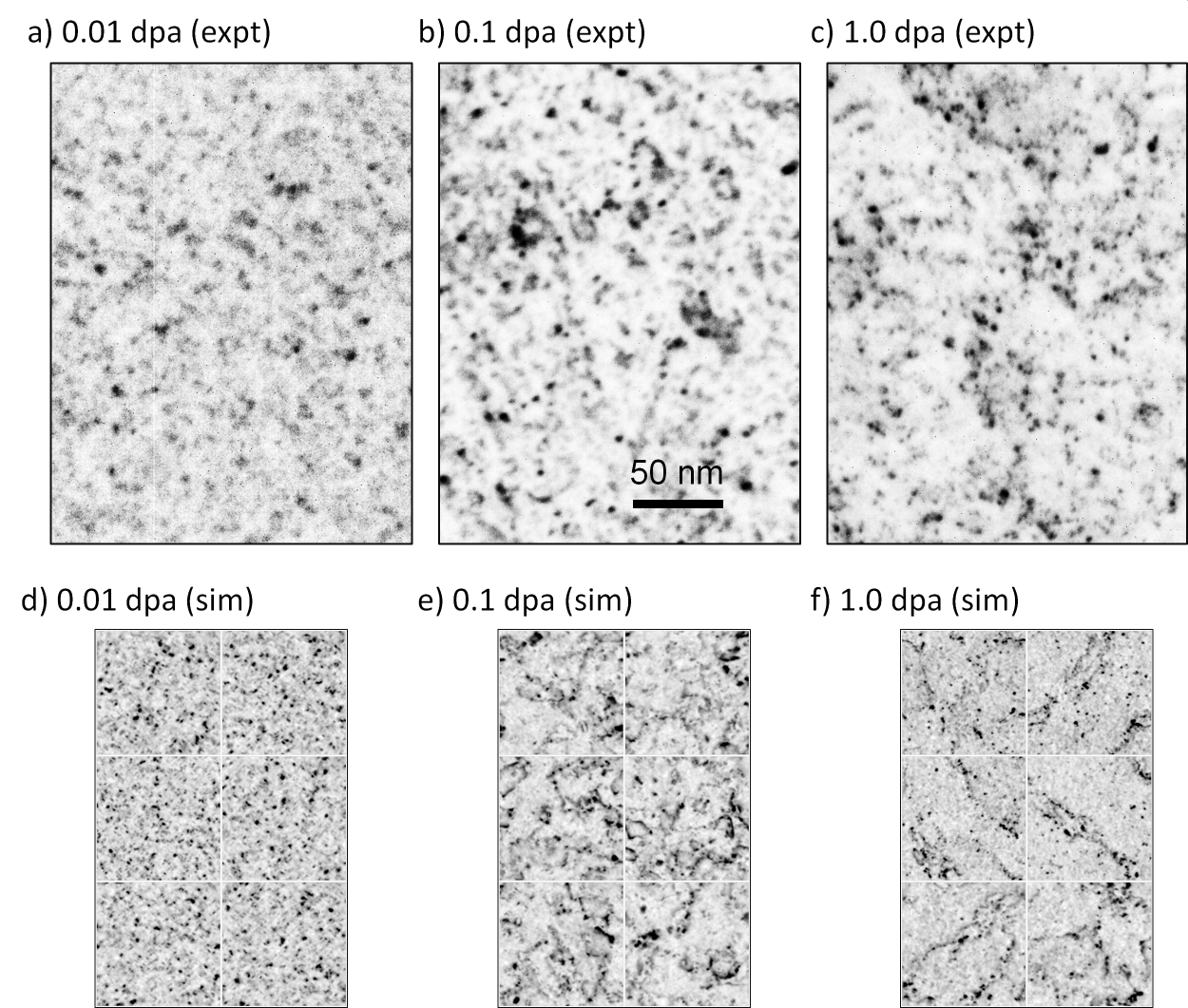}
    \caption{A direct comparison between experimental and simulated weak beam dark field images of irradiated tungsten, shown in inverted contrast with same scale bar and ${\bf g}$=[200] imaging vector. }
\label{fig:comparison_expt_sim}
\end{figure*}

The sequence of images in figure~\ref{fig:sg} demonstrates how the weak beam image offers a correct but incomplete view of the microstructure, and interpreting a high dose microstructure as consisting of quasi-independent dislocation loops from the weak beam image alone can be misleading. Figure~\ref{fig:comparison_expt_sim} makes a direct comparison between experimental weak beam TEM images of ion-irradiated tungsten with the corresponding simulated TEM images. As the experimental images cover a much larger area with a long spatial correlation length compared to the simulations, we artificially increase the apparent simulated area by stitching together six symmetry-related ${\bf g}$-vector - zone axis orientations of the same simulation.

Comparing these image shows the same qualitative behaviour in experiment and simulation. 
At low dose (0.01 dpa) the image is one of isolated black dots, apparently homogeneously distributed.
At intermediate dose (0.1 dpa), the dots show emergent spatial correlation, appearing clustered rather than homogeneous.
At high dose (1.0 dpa), the spatial arrangement is of strings of dots.
With the ground-truth dislocation microstructure, we therefore identify the low dose as dislocation loops, the intermediate as complex dislocation networks, and the high dose as curved sections of dislocation lines. This is the same conclusion as that drawn by Wang {\it et al.}~\cite{Wang_Acta2023} looking at dynamical two-beam images. 

We note that the spot size in our simulations is smaller than seen in experiment.
As noted above, this is partly potential dependent, though it may also be related the fact that loops at room temperature will have non-zero mobility, even when self-trapped by elastic interactions~\cite{Mason_JPCM2014}, and so be able to coalesce to some extent at timescales beyond those accessible with MD.

\begin{figure}
    \centering
    \includegraphics[width=0.9\linewidth]{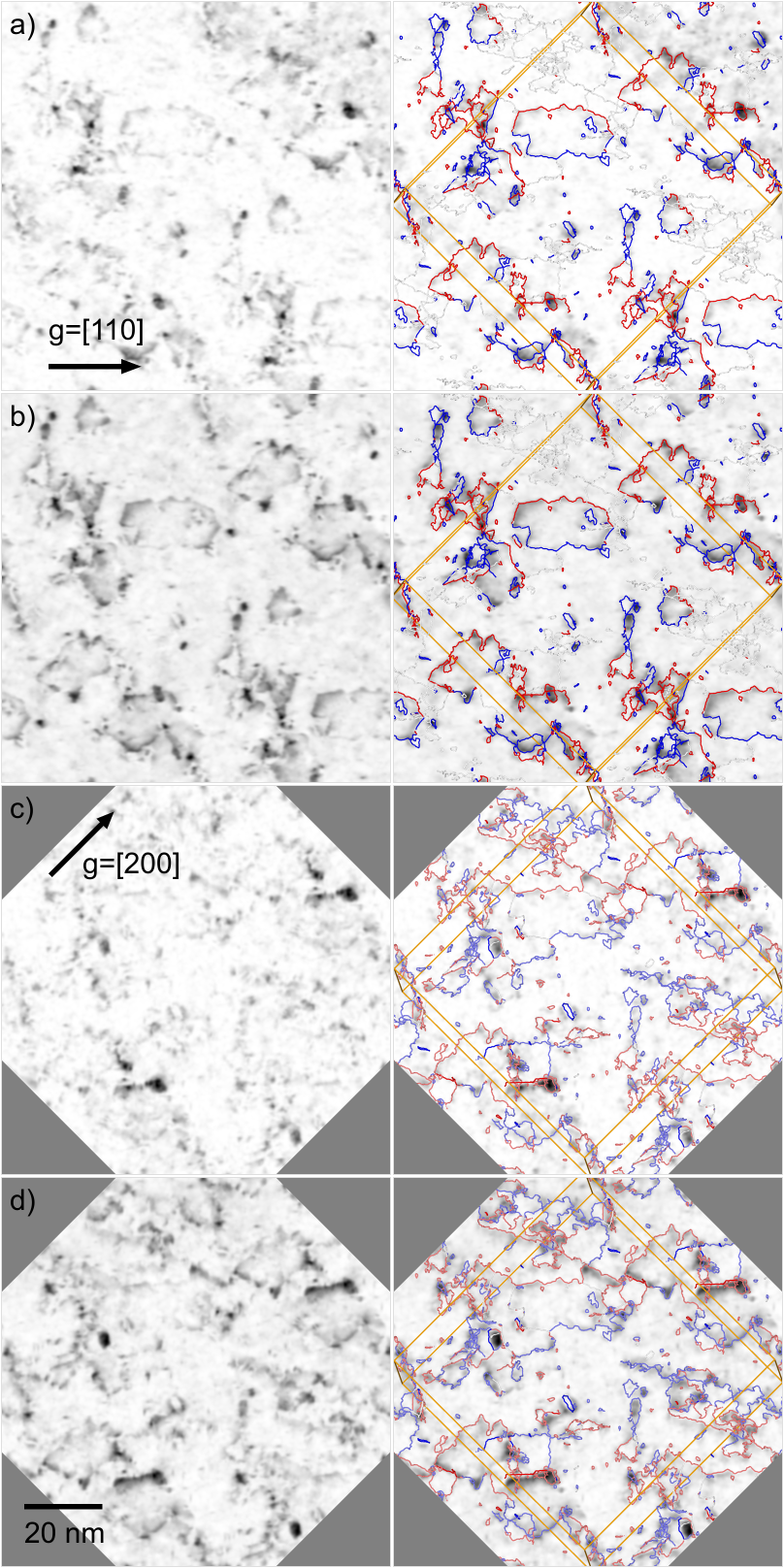}
    \caption{A snapshot of a massively overlapping cascade simulations at 0.10 dpa, with zone axis close to the [001] direction. Simulated TEM with g-vectors ${\mathbf g}=[110]$, $n_{\bf g}$ = 6.25 in images a) and b) and ${\mathbf g}=[200]$, $n_{\bf g}$ = 4.0 in c) and d). Images b) and d) use the precession technique. Images are shown in reverse contrast. Right: dislocation network coloured by $\bf{g}\cdot\bf{b}$ ( red $\bf{g}\cdot\bf{b} = +1$, white $\bf{g}\cdot\bf{b} = 0$, blue $\bf{g}\cdot\bf{b} = -1$ ).}
    \label{fig:g_vs_prec}
\end{figure}

A slightly better conventional TEM image can be generated by using the precession method of Haley {\it et al.}~\cite{Haley_ActaMat2017}.
This is a microscopy technique similar in intent to the convergent weak-beam method of Prokhodtseva {\it et al.}~\cite{Prokhodtseva_ActaMat2013}, in that both methods sample a small range of deviation parameters $s_g$ by rocking the incident beam angle over a small angle.
Where Prokhodtseva used a linear tilt, Haley precesses the beam in a circle. Both suggest averaging over order 10 images with the angle chosen so that the diffracted spot remains within the objective aperture.
In simulation we use a tilt angle of 5 mrad and 10 images. Taking more images does not improve the result to the naked eye.
The result of this procedure is shown in figure~\ref{fig:g_vs_prec}.
Two ${\bf g}$-vectors are shown here, both in weak beam conditions, $s_{\bf g} \sim 0.2 \rm{nm}^{-1}$.
For both imaging vectors the contrast oscillations due to variations in strain in the background and the beading seen in dislocation lines,  exaggerated by high deviation parameter, are smoothed out by averaging over several images with varying $s_{\bf g}$~\cite{Cockayne_JMicro1973}.
This has the effect of joining the black spots into more clear lines.
However the $\bf{g}\cdot\bf{b}$ invisibility criterion~\cite{Jenkins2001}  is still clearly in effect, with approximately half the dislocation network visible at each ${\bf g}$-vector.

This observation leads us to conclude that it will be difficult to establish the true extent of a complex dislocation network using TEM, unless care is taken to superimpose the images from a range of $\bf{g}$-vectors. A dramatic example of this can be seen in Ref~\cite{ElAtwani_Acta2018}, where elastic energy minimisation leads to the majority of large loops having the same Burgers vector at high dose.

\section{Conclusions}
\label{sec:conclusion}

In this paper we have described a simple transformation of the Howie-Whelan equations which enables their quick and efficient evaluation over arbitrarily complex atomic configurations without the need to find displacement or strain fields.
This is done by constructing the complex phase $x({\bf{r}}) = \exp( i \bf{g}  \cdot \bf{r} )$ at atom sites, and interpolating to form a continuous field. 

Partly for didactic purposes, we showed how established results in expert TEM lab use transfer to simulated TEM image generation.
We showed how changing the deviation parameter highlights different parts of the strain field, and how each images produced show a strong spatial correlation to the ground truth dislocation structure, and how our simulated microstructure is a good qualitative fit to both dynamical and weak beam images in the literature.
We conclude that the generation and evolution of our low-temperature high dose microstructure in simulation is a match to the description of network formation in Wang {\it et al.}~\cite{Wang_Acta2023}, and does not require the diffusion of mobile loops.
Importantly here we were able to demonstrate unambiguously, by comparing to ground-truth dislocation microstructure, how the network of dislocations produced by high dose irradiation is almost invisible in weak-beam dark-field imaging due to a combination of the suppression of the image intensity of small strain fields and $\bf{g}\cdot\bf{b}$ invisibility.

We conclude that attempts to model the irradiated microstructure with simple dislocation loop objects, such as used in object kinetic Monte Carlo~\cite{Domain_JNM2004,Becquart_JNM2009,MartinBragado_CPC2013} or cluster dynamics~\cite{Marian_JNM2011}, must fail at the high-dose limit where the formation of a complex dislocation network is driven by fluctuating elastic stresses rather than diffusing point-like objects. 
Conversely, attempts to characterize high-dose microstructure in terms of a size-frequency distribution of dislocation loops will also be misleading, without first demonstrating that a complex dislocation network has not formed.

\section{Acknowledgements}

The authors would like to thank Hongbing Yu and Chris Grovenor for helpful insights and Andrew Warwick for code testing.

This work has been carried out within the framework of the EUROfusion Consortium, funded by the European Union via the Euratom Research and Training Programme (Grant Agreement No. 101052200 - EUROfusion), and by the RCUK Energy Programme, Grant No. EP/W006839/1. To obtain further information on the data and models underlying the paper please contact PublicationsManager@ukaea.uk. The views and opinions expressed herein do not necessarily reflect those of the European Commission. The authors acknowledge the use of the Cambridge Service for Data Driven Discovery (CSD3) and associated support services provided by the University of Cambridge Research Computing Services \cite{csd3} that assisted the completion of this study.

\section{Data availability}
\emph{The code used to generate the dynamical two-beam images will be made public on acceptance of this paper.}

\section{Author Contributions}
{\footnotesize
S.L.D. and D.R.M. developed the concept and the equations used to solve for simulated TEM images. D.R.M. implemented the code and ran the image simulations. M.B. ran the high dose microstructure simulations. J.H. and E.P. ensured the simulation methodology matched that of expert TEM use. F.H and G.H. provided high dose experimental TEM images.
}

\bibliography{references}

\section{Supplementary Material}

\subsection{Validation by comparison to previous work}
\label{sec:validation}

We validate the formulation of the Howie-Whelan equations based on atomic positions only (equations \ref{eqn:two_beam3}) by comparing to the results from an existing TEM simulation code, TEMACI\cite{Zhou_PhilMag_2006} based on analytic formulae for the displacement field in the elastic limit.
We consider four typical irradiation-induced defects which image brightly in weak-beam dark-field conditions, and four which are faint.
The bright defects are prismatic interstitial loops with $1/2[111]$ and $[100]$ Burgers vectors, with diameters of 2 nm and 6 nm, imaged with $\mathbf{g}=2\pi/a_0 [200]$, and using $s_{\bf g} = 0.18$nm$^{-1}, \xi_0 = 13.4,\xi_{\bf g} = 36.6$ nm.
The defects were placed at a depth 25 nm in a 50 nm thick foil, and relaxed with fixed zero strain boundary conditions using LAMMPS\cite{LAMMPS} using the MNB empirical tungsten potential\cite{Mason_JPCM2017}. 
The simulation box was $80 \times 80 \times 160$ conventional cubic unit cells (2 M atoms).

\begin{figure}
    \centering
    a)
    \includegraphics[width=0.3\linewidth]{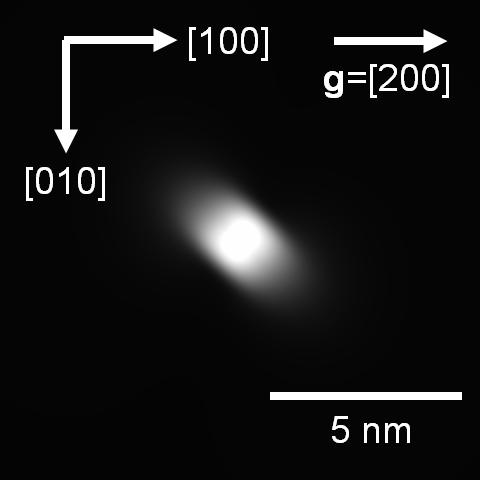}
    b)
    \includegraphics[width=0.3\linewidth]{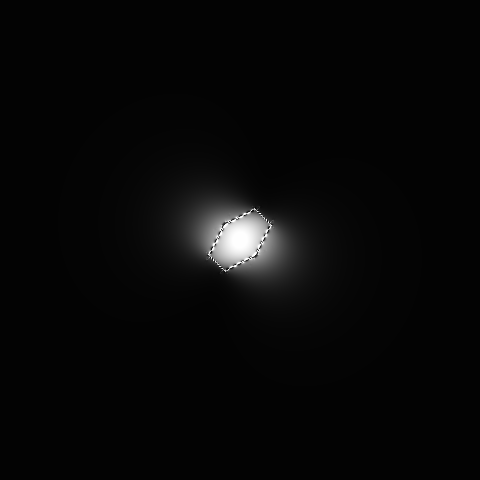}\\
    c)
    \includegraphics[width=0.3\linewidth]{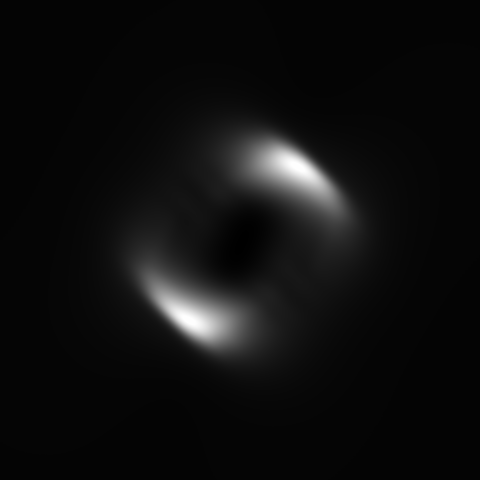}
    d)
    \includegraphics[width=0.3\linewidth]{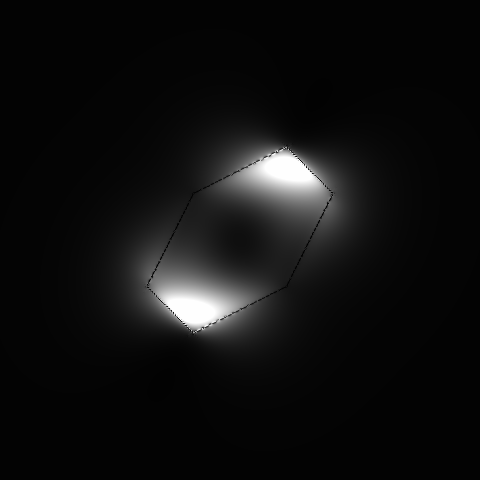}\\
    e)
    \includegraphics[width=0.3\linewidth]{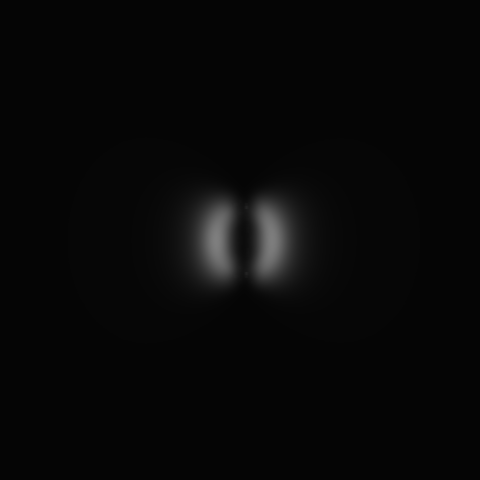}
    f)
    \includegraphics[width=0.3\linewidth]{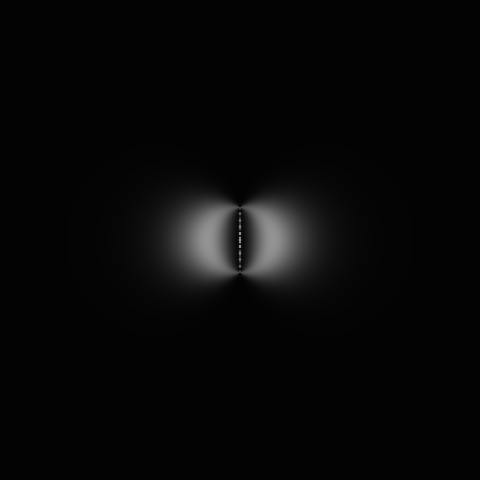}\\
    g)
    \includegraphics[width=0.3\linewidth]{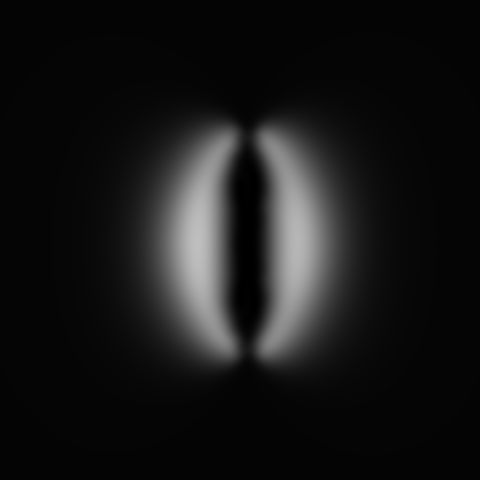}
    h)
    \includegraphics[width=0.3\linewidth]{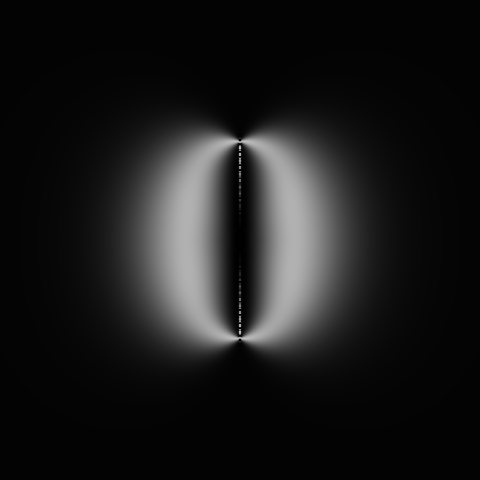}
    \caption{A comparison between simulated TEM images produced from atomistic data (a,c,e,g)  and elastic strain (b,d,f,h).
    Top to bottom: 2 nm 1/2 [111] loop, 6 nm 1/2 [111] loop, 2 nm [100] loop, 6 nm [100] loop at a depth 25 nm in a 50 nm foil.
    Weak beam dark field imaging conditions $s_g=0.1777$ nm$^{-1}$, equivalent to $(\mathbf{g},4g)$ with $\mathbf{g}=[200]$. 
    All images have equal intensity range. 
    }
    \label{fig:comparison_loops1}
\end{figure}

\begin{figure}
    \centering
    a)
    \includegraphics[width=0.3\linewidth]{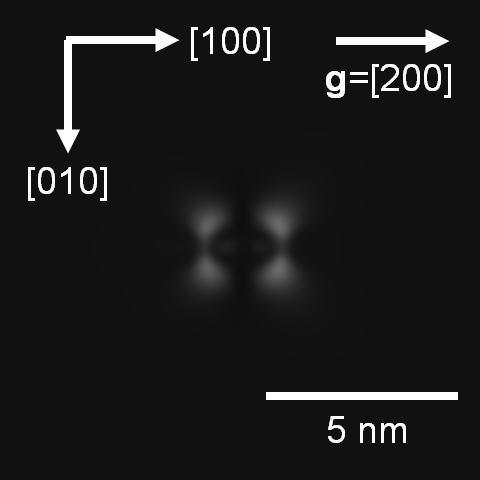}
    b)
    \includegraphics[width=0.3\linewidth]{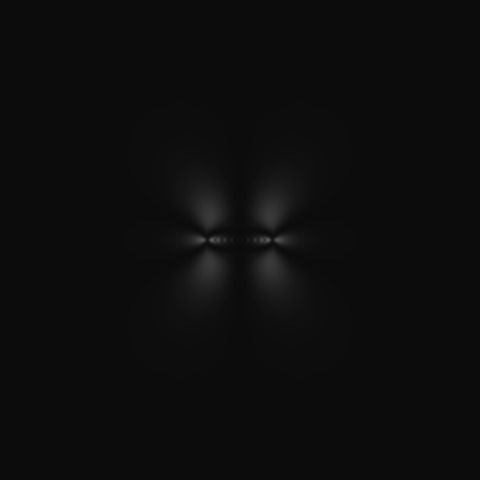}\\
   c)
    \includegraphics[width=0.3\linewidth]{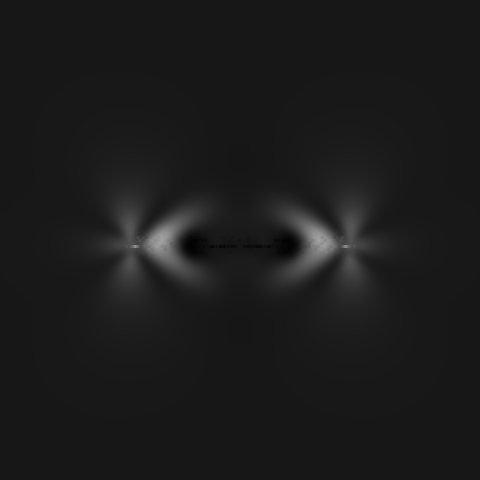}
    d)
    \includegraphics[width=0.3\linewidth]{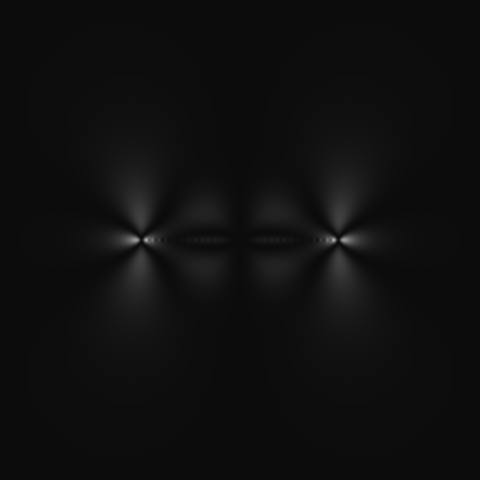}\\
    e)
    \includegraphics[width=0.3\linewidth]{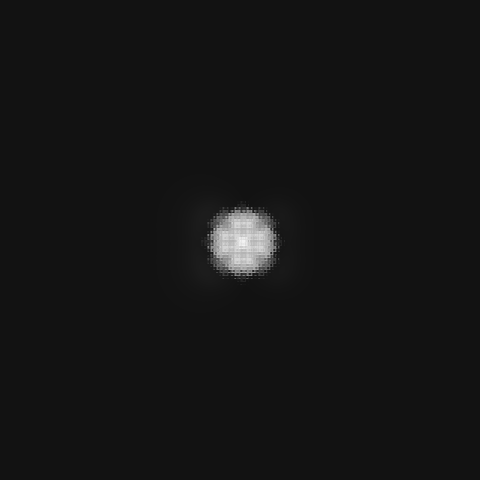}
    f)
    \includegraphics[width=0.3\linewidth]{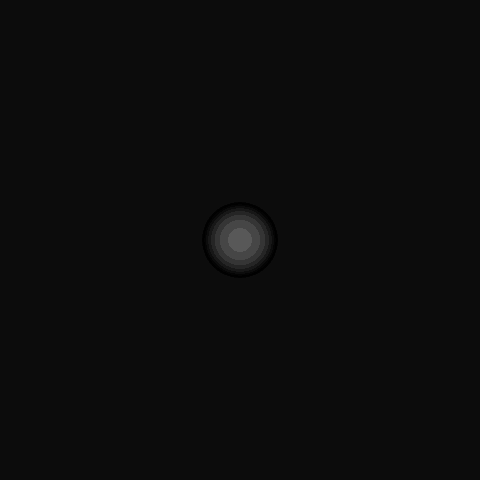}\\
   g)
    \includegraphics[width=0.3\linewidth]{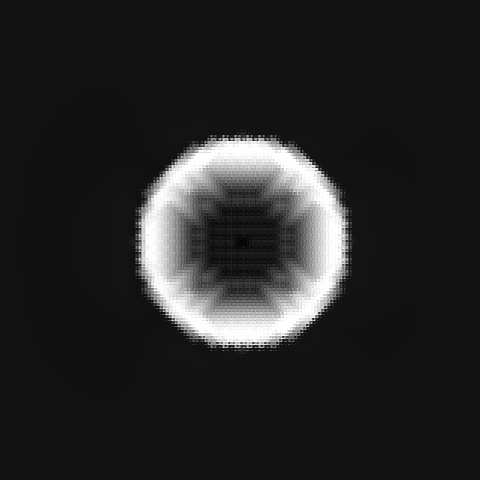}
    h)
    \includegraphics[width=0.3\linewidth]{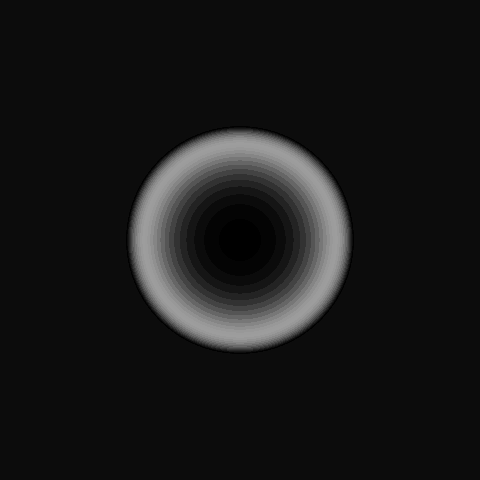}
    \caption{A comparison between simulated TEM images produced from atomistic data (a,c,e,g)  and elastic strain (b,d,f,h).
    Top to bottom: 2 nm  [010] loop, 6 nm  [010] loop, 2 nm void, 6 nm void at a depth 25 nm in a 50 nm foil.
    Weak beam dark field imaging conditions $s_g=0.1777$ nm$^{-1}$, equivalent to $(\mathbf{g},4g)$ with $\mathbf{g}=[200]$. 
    All images have equal intensity range, with maximum intensity level set to one quarter that of figure \ref{fig:comparison_loops1}. Note the relaxed atomistic simulation cell has double the linear dimension of the image, to minimise periodic boundary effects. }
    \label{fig:comparison_loops2}
\end{figure}

The faintly imaging defects were prismatic interstitial loops with $[010]$ Burgers vectors, so that $\mathbf{g}\cdot\mathbf{b} = 0$, and voids. Again these had diameters 2 nm and 6  nm and relaxed as above.
The bright defect simulated TEM images are shown in figure \ref{fig:comparison_loops1}, and the faint in figure \ref{fig:comparison_loops2}.
For the bright defects the saturated white level is set to $|\phi _{\bf g}|^2 = 0.15$, for the faint defects the white level is at $|\phi_{\bf g}|^2 = 0.0625$.

The TEMACI images technically solve the related Howie-Basinski equations\cite{Howie_PhilMagA_1968}. In the columnar approximation, and with two beams only, the Howie-Basinski equations reduce to the Howie-Whelan equations that we use here. The elastic strain fields used were computed analytically by approximating the loops as a hexagon of dislocation line segments.

It is clear by eye that the two sets of images are very similar, clearly showing the same extent and symmetries. To make a more systematic comparison, we find the correlation between the images.
We can find the similarity between two images, whose intensities at 2d position $\mathbf{x}$ are given by $f(\mathbf{x})$ and $g(\mathbf{x})$, using a simple correlation function, $c$, defined by
\begin{equation}
    \label{eqn:correlation_function}
    c = \frac{\int \left( f(\mathbf{x} ) - \bar{f} \right) \left( g(\mathbf{x} ) - \bar{g} \right) \mathrm{d}^2\mathbf{x} } { \sqrt{ \int \left( f(\mathbf{x}) - \bar{f} \right)^2  \mathrm{d}^2\mathbf{x} \, \int \left( g(\mathbf{x}) - \bar{g} \right)^2  \mathrm{d}^2\mathbf{x}  } },
\end{equation}
where $\bar{f},\bar{g}$ are the average intensities in each image respectively.
This correlation function 
is invariant to affine transformations of the intensity ( brightness and contrast shifts ) and weakly dependent on noise providing the signal-to-noise ratio 
is large.
A correlation value of one means that two images have identical (relative) spatially varying intensities, a value of minus one means they form a pair of negatives.

Table \ref{tab:correlation_atomic_strain} shows the correlation factors between the eight images. If we read across each row looking for the closest match between an atomistic TEM image and a strain field TEM image, we find a very highly correlated like-for-like match in most cases, with the exception being that the atomistic image for the `invisible' $\mathbf{b}=[010]$ loops could still be mistaken for a small $\mathbf{b}=[100]$ loop.

\begin{table*}[htb!]
    \centering
    \begin{tabular}{l|cccccccc}
                    &   \multicolumn{8}{c}{TEMACI strain formulae}      \\
        atomic      &   $[111]$(2nm)  &  $[111]$(6nm)  &  $[100]$(2nm)  &  $[100]$(6nm)  &  $[010]$(2nm)  &  $[010]$(6nm)   &   void (2nm)   &   void (6nm)  \\
        \hline
$[111]$(2nm)    & \bf{0.91}    &    0.26    &    0.83    &    0.55    &    0.51    &    0.30    &    0.71    &    0.23     \\ 
$[111]$(6nm)    & 0.13    &    \bf{0.94}    &    0.40    &    0.58    &    0.53    &    0.15    &    0.00    &    0.69     \\
$[100]$(2nm)    & 0.72    &    0.46    &    \bf{0.90}    &    0.74    &    0.71    &    0.34    &    0.72    &    0.00     \\
$[100]$(6nm)    & 0.51    &    0.57    &    0.82    &    \bf{0.94}    &    0.75    &    0.38    &    0.17    &    0.62     \\
$[010]$(2nm)    & 0.32    &    0.37    &    \bf{0.74}    &    0.65    &    0.74    &    0.24    &    0.04    &    0.49     \\
$[010]$(6nm)    & 0.25    &    0.21    &    \bf{0.54}    &    0.52    &    0.42    &    0.50    &    0.04    &    0.31     \\
void (2nm)      & 0.87    &    0.04    &    0.35    &    0.06    &    0.18    &    0.04    &    \bf{0.92}    &    -0.05    \\
void (6nm)      & 0.07    &    0.66    &    0.22    &    0.77    &    0.30    &    0.45    &    -0.03    &    \bf{0.97}      
    \end{tabular}
    \caption{Image correlation factors between simulated TEM images generated using atomic position data (figures \ref{fig:comparison_loops1},\ref{fig:comparison_loops2} a,c,e,g) and strain formulae using TEMACI (same figures b,d,f,h). For each row ( each atomic figure ) the best correlated TEMACI figure is highlighted.}
    \label{tab:correlation_atomic_strain}
\end{table*}

\subsection{Finding good two-beam diffraction conditions}
\label{sec:two_beam}

In this section we derive equations required for finding good dynamical two beam conditions.
Start with an incident high energy electron beam with wave vector given by 
    \begin{equation}
         \label{eqn:incident_wavevector}
        \mathbf{k} = \frac{m v}{\hbar} \hat{\mathbf{k}},
    \end{equation}
where $v$ is the velocity of the electrons, and $\hat{\mathbf{k}}$ a unit vector in the direction of the beam.  

We have a set of atoms in general position defining the foil. 
For good diffraction contrast, we hope most of the atoms are in a lattice, so we encode the conventional unit cell of the atomic lattice as the matrix $\mathbf{C}$, where the component $C_{ij}$ is the $i^{th}$ Cartesian component of the $j^{th}$ lattice vector in lab frame. For a cubic unit cell, oriented with a zone axis along $[001]$, this matrix might be simply $\mathbf{C} = a_0 \mathbf{I}$, where $\mathbf{I}$ is the $3\times3$ identity matrix, but more generally the columns of $\mathbf{C}$ are the three conventional lattice vectors $\mathbf{c}_1,\mathbf{c}_2$ and $\mathbf{c}_3$, ie
    \begin{equation}
        \mathbf{C} =   \left( \begin{array}{ccc}  
                            \mathbf{c}_1   &  \mathbf{c}_2    &   \mathbf{c}_3   \\
                            \vert       &  \vert        &   \vert   \\
                            \vert       &  \vert        &   \vert   
                        \end{array} \right).
    \end{equation}
$\mathbf{C}$ will be determined by the positions of the atoms in the input file, which may have any orientation or strains present, and not by the bounding simulation box.

Note that this equates the coordinate frame of the input file with the lab frame, while the crystal frame is allowed to vary.

The reciprocal lattice vectors by convention are given by $\mathbf{b}_1 = 2 \pi \, \mathbf{c}_2 \times \mathbf{c}_3 / ( \mathbf{c}_1 \cdot (\mathbf{c}_2 \times \mathbf{c}_3) )$ etc, which can be encoded in the matrix $\mathbf{B} = 2 \pi \mathbf{C}^{-1}$, so that the $[hkl]$ reciprocal lattice vector is a vector in the crystal frame given by
    \begin{equation}
        \label{eqn:reciprocal_lattice_vectors}
        \mathbf{g}_{hkl} = \mathbf{B}^T \left(
        \begin{array}{c} 
            h   \\
            k   \\
            l
        \end{array} \right).
    \end{equation}

If we solve the dynamical two beam (Howie-Whelan) equations using the imaging reciprocal vector $\mathbf{g}$, we can define the (energy) deviation parameter
    \begin{equation}
        \label{eqn:deviation_parameter}
        \epsilon_{\mathbf g} = \frac{\hbar^2 ({\mathbf k}+{\mathbf g})^2}{2m}-\frac{\hbar^2 {\mathbf k}^2}{2m} = \frac{\hbar^2 (2\mathbf{k}+\mathbf{g}) \cdot \mathbf{g}} {2m}.
    \end{equation}
    
Combining equations \ref{eqn:incident_wavevector},\ref{eqn:reciprocal_lattice_vectors},\ref{eqn:deviation_parameter}, we find the Bragg condition is met at the $[hkl]$ reciprocal lattice vector when
    \begin{equation}
        \label{eqn:Bragg_condition}
        \frac{ 2 m v }{\hbar}\,
        \hat{\mathbf{k}} \cdot
         \mathbf{B}^T \left(
        \begin{array}{c} 
            h   \\
            k   \\
            l
        \end{array} \right)  +
         \left| \mathbf{B}^T \left(
        \begin{array}{c} 
            h   \\
            k   \\
            l
        \end{array} \right) \right|^2 = 0.
    \end{equation}
This condition is trivially met for $[hkl]=[000]$, and can be solved for other reciprocal vectors by rotating the incident beam vector $\hat{\mathbf{k}}$.
We will look for a rotation where this condition is met for a particular $[hkl]$, but not met for other reciprocal lattice vectors $[h'k'l']$.


We start by defining an orthonormal basis set $\hat{\mathbf{x}},\hat{\mathbf{y}},\hat{\mathbf{z}}$. 
The user selects a zone axis direction, $[uvw]$, with a direction in the lab frame given by $\hat{\mathbf{z}} \sim \mathbf{C} [uvw]^T$.
We choose a second unit vector $\hat{\mathbf{y}}$ to have no projection along the desired $g$-vector direction. 
To align the crystal with the incident beam, we therefore first rotate the crystal by $\mathbf{U}$ given by
    \begin{equation}
        \mathbf{U} =   \left( \begin{array}{ccc}  
                            \hat{\mathbf{x}}     &  \horz        &  \horz   \\
                            \hat{\mathbf{y}}     &  \horz        &   \horz   \\
                            \hat{\mathbf{z}}     &  \horz        &   \horz   
                        \end{array} \right),
    \end{equation}
and then use a virtual `tilt stage' to fine-tune the orientation of the crystal through a rotation matrix $\mathbf{R}$ in order to get a good two-beam diffraction pattern.
Under these two rotations a vector $\mathbf{x}$ rotates to $\mathbf{R} \mathbf{U} \mathbf{x}$.
The $\mathbf{g}-$vector rotates to $\mathbf{g}' = \mathbf{R} \mathbf{U} \mathbf{g}$ so that for an atom at position $\mathbf{x}$, we see $\mathbf{g}'\cdot\mathbf{x}' = \mathbf{g}\cdot\mathbf{x}$, as required.

If our tilt stage first rotates by $\theta$ about the x-axis, then by $\psi$ about the y-axis, the rotation matrix describing the transformation is 
    \begin{equation}
        \mathbf{R} = \left(
        \begin{array}{ccc} 
             \cos \psi & \sin \theta \sin \psi & \cos \theta \sin \psi  \\
             0 &   \cos \theta   & -\sin \theta \\
             -\sin \psi  & \sin \theta \cos \psi   & \cos \theta \cos \psi 
        \end{array} \right).
    \end{equation}    

The Bragg condition, equation \ref{eqn:Bragg_condition}, after rotation reads
    $\hat{\mathbf{k}}\cdot \mathbf{g}'  = \alpha$, where $\alpha = -\hbar  | \mathbf{g} |^2 /(2mv)$ is a constant, so if the components of $\mathbf{g}$ are
    \begin{equation}
        \left( \begin{array}{c}
            g_1 \\
            g_2 \\
            g_3 
          \end{array} \right)
          = \mathbf{U} \mathbf{B}^T 
          \left( \begin{array}{c}
            h \\
            k \\
            l 
          \end{array} \right),
    \end{equation}
this gives the vector equation 
    \begin{equation}
        \label{eqn:SolveForPsi}
        \left(  \begin{array}{c} 
            - \sin \psi             \\
            \cos \psi \sin \theta   \\
            \cos \psi \cos \theta
        \end{array} \right) 
            \cdot 
        \left(  \begin{array}{c} 
            g_1 \\
            g_2 \\
            g_3
        \end{array} \right) 
        = \alpha
    \end{equation}
    

Equation \ref{eqn:SolveForPsi} has solutions $\{\psi\}$ for each choice of $\theta$, given by
    \begin{eqnarray}
        \psi &=& \pm \rm{acos} \left[
         u \pm v
        \right] +  2 n \pi \quad   \quad (n \in \mathcal Z),    \nonumber\\
        u &=& \frac{ \alpha ( g_2 \sin \theta + g_3 \cos \theta ) } { g_1^2 + ( g_2 \sin \theta + g_3 \cos \theta )^2 } \nonumber \\
        v &=& \frac{ g_1 \sqrt{  g_1^2 + ( g_2 \sin \theta + g_3 \cos \theta )^2 - \alpha^2 } } { g_1^2 +  ( g_2 \sin \theta + g_3 \cos \theta )^2 } .
    \end{eqnarray}
We therefore need an algorithm to decide which pair $\{\psi,\theta\}$ represents best diffraction conditions.

For the perfect crystalline foil thickness $L$, the equations to solve for the propagation of incident and diffracted electron beams (equation \ref{eqn:two_beam3}) reduce to:
\begin{equation}    
    \label{eqn:two_beam6}
{\partial \over \partial z}
\left( \begin{array}{c}
    \phi_0(z) \\
    \phi_{\bf{g}} (z)
\end{array} \right)
=
i \pi 
\left( \begin{array}{c c}
    1/\xi_0     &   1/\xi_{\bf{g}}      \\
    1/\xi_{\bf{g}}     &   1/\xi_0 + 2 s_{\bf{g}}    
\end{array} \right) 
\left( \begin{array}{c}
    \phi_0(z) \\
    \phi_{\bf{g}} (z)
\end{array} \right),  
\end{equation}
This can be solved analytically with the boundary conditions $\phi_0(0) = 1, \phi_{\bf{g}} = 0$, to give the solution at the back face of the foil, $z=L$,
    \begin{eqnarray}
        \label{eqn:propagation_perfect_foil}
        \left|\phi_0(L)\right|^2 &=& \frac{s_{\bf{g}}^2 + \xi_{\bf{g}}^{-2} \cos^2\left[ \pi \sqrt{ s_{\bf{g}}^2 + \xi_{\bf{g}}^{-2} } L \right] }{s_{\bf{g}}^2 + \xi_{\bf{g}}^{-2}}      \nonumber\\
        \left|\phi_{\bf{g}}(L)\right|^2 &=& \frac{\xi_{\bf{g}}^{-2} \sin^2\left[ \pi \sqrt{ s_{\bf{g}}^2 + \xi_{\bf{g}}^{-2} } L \right] }{s_{\bf{g}}^2 + \xi_{\bf{g}}^{-2}}.   
    \end{eqnarray}
The foil thickness $L$ is found in the imaging space, defined in section \ref{sec:image_space}.
Note that both $s_{\bf{g}}$ and $L$ depend on the virtual tilt rotation  $\mathbf{R}$, 

We can now define a score for a rotation $\mathbf{R}$, 
    \begin{equation}
        S = \sum_{\mathbf{g}} \left|\phi_{\bf{g}}(L;\mathbf{R})\right|^2,
    \end{equation}
where the sum is over a range of reflections excluding $[000]$ and the requested reflection $[hkl]$.
The best rotation for dynamic two beam imaging has the smallest value $S$, this corresponds to the rotation where unwanted reflections have minimum intensity.
For our score, we consider the set of permitted reflections $[hkl]$ where $h^2 + k^2 + l^2 \le 10^2$.

If we set the $n_g^{\rm{th}}$ diffraction spot, ie $n_g \times [hkl] $ bright, and others faint, we then place the `aperture' at $[hkl]$. (Noting that in the Howie-Whelan approximation the aperture is a delta function. ). Large $n_g$ corresponds to a large deviation parameter $s_g$. 
In this work we have also used fractional values for $n_g$, which are more appropriate in experimental work, as strongly-excited systematic reflections can cause extinction effects in the image if integer values are chosen~\cite{Jenkins2001}.
To find a fractional value, we find the quaternions for the integer values of $n_g$, and spherical interpolate between as appropriate~\cite{Shoemake_SIGGRAPH85}.

\subsection{Mapping atoms to the image space}
\label{sec:image_space}

In this section we consider how the atoms in the input file are placed into an imaging space, whose length and breadth define the extend of the pixels in the output image. This enables us to define the foil thickness.
\begin{figure}
    \centering
    \includegraphics[width=0.9\linewidth]{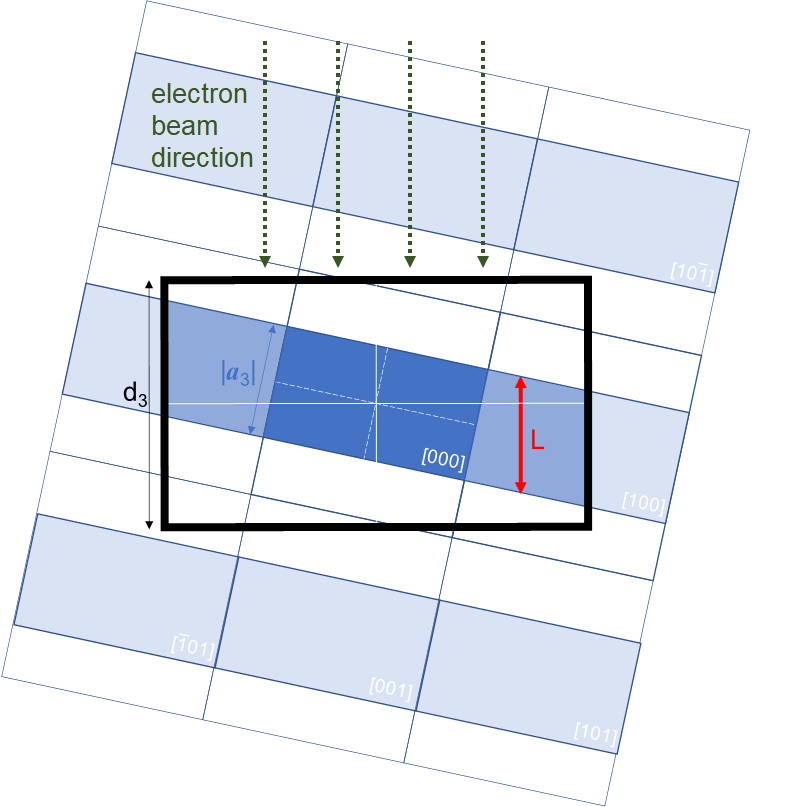}
    \caption{Cartoon illustrating the geometry of the periodic input cell and the imaging space. The input file contains atoms in the $[000]$ periodic replica, illustrated by the dark shaded box. Periodic replicas inside the imaging space ( heavy black line box ) are kept. }
    \label{fig:imagingSpace}
\end{figure}
We start by stating that the atoms are in a periodic supercell whose repeat vectors, of length 100 nm or more, are given by the columns of the matrix $\mathbf{A}$, where
    \begin{equation}
        \mathbf{A} =   \left( \begin{array}{ccc}  
                            \mathbf{a}_1   &  \mathbf{a}_2    &   \mathbf{a}_3   \\
                            \vert       &  \vert        &   \vert   \\
                            \vert       &  \vert        &   \vert   
                        \end{array} \right).
    \end{equation}
If periodicity in one direction is not desired (to represent a foil), and it is not present in the original file, then the user can add padding in that direction. In figure \ref{fig:imagingSpace}, we show an input file with a free surface in one direction, represented by the heavy shaded box in the centre.

After an affine transformation consisting of a rotation $\mathbf{U}$, and translation $\bm{\delta}$, the position of an atom in the $[ijk]$ periodic replica of $\mathbf{A}$ transforms to 
    \begin{equation}
        \label{eqn:transformation}
        \mathbf{x}' = \mathbf{U} \left( \mathbf{x} + \mathbf{A}\left( \begin{array}{c}  
                            i   \\
                            j   \\
                            k
                        \end{array} \right) \right) + \bm{\delta}.
    \end{equation}
The periodic replicas of the atoms in the input file after rotation and translation are shown with light shaded boxes in figure \ref{fig:imagingSpace}.
The atom replicas are placed in in \emph{imaging space}, defined as a new orthorhombic box, whose lattice vectors are given by the columns of the matrix $\mathbf{D}$, given by
    \begin{equation}
        \mathbf{D} =   \left( \begin{array}{ccc}  
                            d_1   &  0      &   0   \\
                            0     &   d_2   &   0   \\
                            0     &  0      &   d_3   
                        \end{array} \right),
    \end{equation}
where $d_1$ and $d_2$ are the size of the output image (in nm) and $d_3$ is the integration depth total, chosen by the user.
The original periodic boundaries across replicas of $\mathbf{A}$ will be respected within the transformed cell, but there is no periodicity across replicas of $\mathbf{D}$-- rather this transformed cell should be viewed as a cuboidal crystallite whose sides have lengths $d_1,d_2,d_3$, suspended in vacuum.
The imaging space is represented by the heavy lined box in figure \ref{fig:imagingSpace}, with its 3-axis aligned with the electron beam.
    
The vector offset for the affine transformation in equation \ref{eqn:transformation}, $\bm{\delta}$, is given by 
    \begin{equation}
        \label{eqn:translation}
        \bm{\delta} =  
        \mathbf{D}\left( \begin{array}{c}  
                            \half   \\
                            \half    \\
                            \half 
                        \end{array} \right) 
                - \mathbf{U} \mathbf{A}\left( \begin{array}{c}  
                            \half   \\
                            \half    \\
                            \half 
                        \end{array} \right),
    \end{equation}
so that an atom in the centre of the first periodic replica of the input file is transformed to the centre of the image.

Succinctly, the $[ijk]$ replica of an atom at $\mathbf{x}$ in the input file is within the imaging space if for each Cartesian component $\alpha \in\{1,2,3\}$,
    \begin{equation}
        \mathrm{int}\left( \mathbf{U}\left( \mathbf{x}+\mathbf{A} \left[ i-\half,j-\half,k-\half \right]^T \right)_{\alpha}/d_{\alpha} + \half\right) = 0.
\end{equation}
In figure \ref{fig:imagingSpace}, some atoms from the central replica $[ijk]=[000]$, and its neighbours $[\bar{1}00]$,$[100]$ are included in the imaging space.

The thickness of the foil, $L$, can then be determined from the positions of the atoms in the imaging space, as shown in figure \ref{fig:imagingSpace}.
Note that without the periodic replicas, the edges of the image do not have the same thickness: periodic replicas are required to see the full information contained in the atomic input file.
We find the thickness by conceptually dividing the atoms in $\mathbf{D}$ into columns aligned with the 3-axis, and finding the maximum and minimum height of atoms in that column. The foil thickness for the column is given by the difference plus half a lattice parameter, ie $a_0/2$.
The mean foil thickness $L$ is the average thickness of the columns, and the mean normal $\hat{\mathbf{n}}$ to the surface from the average spatial variation of the maxima and of the minima. Under a small tilt stage rotation $\mathbf{R}$, the foil thickness changes to 
    \begin{equation}
        L' = \frac{ [001] \hat{\mathbf{n}} }{ [001] \mathbf{R} \hat{\mathbf{n}}   } L.
    \end{equation}

\end{document}